\documentclass{article}
\usepackage[utf8]{inputenc}
\usepackage[T1]{fontenc}
\usepackage{spconf,amsmath,graphicx,booktabs,amssymb,url,microtype,hyperref}
\usepackage{xcolor}
\usepackage{tikz}
\usetikzlibrary{positioning,calc}

\title{SPHERE: A Multi-Attribute No-Reference Metric for Binaural Speech Quality and Spatial Fidelity}
\name{Bala Manikya Prasad Puram, Chandan K A Reddy, Artem Dementyev}

\address{
  Google LLC \\
  Emails: pbmprasad@google.com, chandanka@google.com, artemd@google.com
}

\begin{document}
\ninept
\setlength{\abovedisplayskip}{3pt}
\setlength{\belowdisplayskip}{3pt}

\maketitle
\begin{abstract}
In this paper, we introduce SPHERE (Spatial and Perceptual Human Evaluation and Rating Estimator), a multidimensional no-reference perceptual evaluation framework for binaural speech. SPHERE defines nine distinct perceptual and spatial attributes that characterize the immersive listening experience. We detail the crowdsourcing protocol and evaluation template used to collect ground-truth human ratings. To demonstrate the tractability of automating this framework, we evaluate six baseline configurations across single-branch and Siamese architectures utilizing Mel-spectrograms, complex STFTs, and pretrained spatial embeddings (PhaseCoder). Experimental results demonstrate that these baseline estimators achieve strong correlation with human subjective ratings, proving the feasibility of automated multidimensional spatial audio quality assessment.
\end{abstract}
\section{INTRODUCTION}

The proliferation of advanced hearables, smart glasses, and augmented reality (AR) devices has made spatial audio rendering and environmental noise suppression critical components of the modern acoustic experience \cite{valin2024ar, doclo2010binaural}. In these binaural systems, preserving precise interaural time and level differences (ITD and ILD) is essential for maintaining accurate sound localization and spatial immersion. However, aggressive non-linear signal processing—such as applying dual-channel neural noise suppression—frequently distorts these delicate spatial cues, leading to unnatural source wandering, localized artifacts, and increased listening effort \cite{doclo2010binaural, schadler2020binaural}.

While traditional objective metrics like PESQ \cite{pesq}, POLQA \cite{polqa}, and ViSQOL \cite{visqol} are highly effective for telephony speech, they are strictly single-channel and fail to capture spatial cues like sound localization and orientation. Recent deep-learning-based no-reference (NR) models, such as NISQA \cite{nisqa}, DNSMOS \cite{dnsmos_p835}, and SESQA \cite{sesqa}, successfully predict multi-dimensional speech quality but remain blind to spatial degradation. Conversely, multichannel metrics such as PEAQ (ITU-R BS.1387) \cite{peaq} and AMBIQUAL \cite{ambiqual} evaluate perceptual similarity in ambisonics and spatial music. However, they require clean, full reference signals and do not assess speech-specific dimensions like listening effort, speech distortion, or neural voice enhancement artifacts.

To address these complexities, we developed a subjective testing template capturing nine distinct perceptual attributes. Built on established standards, it evaluates core quality and effort (ITU-T P.835 \cite{p835}, P.810 \cite{ITUTP810}), spatial immersion (ITU-T P.830 \cite{ITUTP830}), and annoyance factors like artifacts and spatial wandering (ITU-T P.811/Supplement 29 \cite{ITUT_P811}). To automate this evaluation, we introduce SPHERE, a no-reference objective framework predicting these nine attributes from stereo inputs. We evaluate six baseline configurations across single-branch and Siamese architectures utilizing Mel-spectrograms, complex STFTs, and pretrained spatial embeddings (PhaseCoder \cite{dementyev2026phasecoder}). Experimental results demonstrate strong correlation with subjective human ratings, providing a foundational tool for evaluating binaural speech enhancement.

\section{Subjective Evaluation Framework for Binaural Speech quality}
\label{sec:subjective_framework}

\subsection{Crowdsourcing Protocol and Hardware Pre-filtering}
To collect reliable ground-truth perceptual ratings across large-scale binaural audio, we established a crowdsourced listening test protocol following the principles of ITU-T Recommendation P.808 \cite{p808}. Unlike monaural telephony evaluation, binaural spatial audio assessment is uniquely sensitive to listener hardware variability, as acoustic crosstalk from loudspeakers completely collapses interaural time and level differences (ITD/ILD). 

To ensure measurement integrity, the protocol enforces a strict multi-tier pre-filtering procedure:
(1) \textbf{Hardware Pre-filter \& Exclusion}: Prior to accessing tests, raters declare their playback equipment category (wired over-ear, on-ear, in-ear earbuds, or built-in speakers). Laptop, phone, and external speakers are strictly disqualified. Hardware model details and Bluetooth/ANC states are logged to verify that spatial ratings are not contaminated by uncontrolled onboard DSP.
(2) \textbf{Binaural Channel Verification}: To verify true stereo reproduction and correct ear orientation, raters must pass an acoustic screening test consisting of speech and tone tokens played exclusively in either the left or right ear, as well as binaural phase-cancellation checks. Raters failing this check are barred from the study.
(3) \textbf{Volume Calibration}: Listeners adjust computer volume to a comfortable listening level during a standardized digit-in-noise sample and are instructed not to alter volume during evaluation.
(4) \textbf{Listener Qualification \& Anti-Spam}: Raters complete hearing screening questions (disqualifying individuals reporting hearing loss or prior involvement in speech coding) and spam/attention checks (transcribing 3 digits in background noise, and performing an audio math verification). Each test Human Intelligence Task (HIT) contains 10 evaluation clips, 1 hidden pristine gold anchor (rating $<4.0$ disqualifies), and 1 degraded trap clip. Each clip receives a minimum of 5 screened ratings post-filtering ($\text{SEM} \approx 0.35$ per clip); aggregating across conditions ($>100$ clips per condition) reduces the standard error of the mean (SEM) to $< 0.04$, ensuring robust statistical significance for benchmarking.

\subsection{Evaluation Template and Target Attributes}
Ground-truth ratings are collected via a structured 4-part template:
\textbf{Part 1 (Core Quality \& Effort)} evaluates Speech Distortion (MOS-SIG) and Background Intrusiveness (MOS-BAK) on 5-point ACR scales per ITU-T P.835 \cite{p835} (1: Very distorted/intrusive to 5: Not distorted/noticeable), Overall Quality (MOS-OVRL, 1: Bad to 5: Excellent) per P.800/P.835, and Listening Effort (1: Extremely high effort to 5: Very low effort / complete ease) adapted from ITU-T P.810 \cite{ITUTP810}.
\textbf{Part 2 (Spatial Immersion)} assesses Loudness Comfort (P.800, P.830, P.840) to evaluate dynamic compression and Sound Localization (P.830) to evaluate directional orientation accuracy.
\textbf{Part 3 (Annoyance \& Processing Artifacts)} tracks Distracting Artifacts (clicks, pops, frame drops per P.830/P.835), Spatial Stability (unnatural source wandering per ITU-T P.811 / Supp. 29 \cite{ITUT_P811}), and Background Noise Behavior (noise pumping or gating fluctuations per P.835).
\textbf{Part 4 (Conditional Low Score Justification)} provides a structural validation mechanism. Triggered whenever a rater assigns $\text{MOS-OVRL} \le 2$, the rater must identify the primary root cause (excessive background noise, speech distortion, loudness imbalance, or spatial instability). This cross-validates the inter-metric relationships between MOS-SIG, MOS-BAK, and MOS-OVRL, eliminating arbitrary scores and ensuring that the subjective MOS targets represent authentic, interpretable perceptual ground truth.

\begin{table*}[t]
\centering
\caption{The 9 target perceptual attributes, their underlying acoustic/psychoacoustic cues, and their engineering significance in spatial wearables (earbuds, smart glasses, and AR/VR headsets).}
\label{tab:attributes_need}
\resizebox{\textwidth}{!}{
\begin{tabular}{lll}
\toprule
\textbf{Perceptual Metric} & \textbf{Underlying Acoustic/Psychoacoustic Cues} & \textbf{Engineering \& Product Significance in Wearables} \\
\midrule
Speech Quality (MOS-SIG) & Speech distortion, timbral coloration, phase mismatch & Evaluates the impact of voice processing on speaker naturalness and clarity \\
Noise Intrusiveness (MOS-BAK) & Signal-to-Noise Ratio (SNR), background noise levels & Measures the effectiveness of ambient noise suppression for user comfort \\
Overall Quality (MOS-OVRL) & Integrates speech quality, noise suppression, and spatial naturalness & Primary metric for product benchmarking and competitor comparison \\
Listening Effort        & Intelligibility, cognitive fatigue, acoustic clarity & Critical KPI for conversation boost and hearing assistance features \\
Loudness Comfort        & Volume appropriateness, dynamic range compression & Evaluates comfort of automatic volume leveling under varying environments \\
Sound Localization      & Interaural Time/Level Differences (ITD/ILD), HRTFs & Ensures spatial rendering preserves accurate sound source direction \\
Spatial Stability       & Angular tracking stability, ITD/ILD drift over time & Detects unnatural wandering of voices due to head-tracking/dynamic-steering jitter \\
Distracting Artifacts   & Clicks, pops, frame drops, high-frequency digital glitches & Monitors low-latency DSP processing errors and codec-level distortions \\
Background Noise Behavior & Noise pumping, gating fluctuations, residual noise stability & Identifies distracting noise modulation artifacts caused by dynamic VADs \\
\bottomrule
\end{tabular}
\vspace{-0.3cm}
}
\end{table*}

\section{Datasets}
\label{sec:datasets}

To develop a robust no-reference perceptual quality metric for binaural speech, we compiled a large-scale database of binaural audio. This section describes the dataset compilation, the crowdsourcing perceptual labeling methodology, and the subsequent data augmentation and splitting protocols for model training.

\subsection{Dataset Collection and Labeling}

The core database consists of approximately 9,000 binaural audio clips of 10 to 20 seconds duration, designed to evaluate earbud playback and voice enhancement performance. The audio clips comprise: (1) \textbf{Real-time captures} recorded directly from commercial earbuds during active digital signal processing (DSP); (2) \textbf{In-lab HATS recordings} using a Head and Torso Simulator (HATS) in controlled acoustic chambers with varying speaker distance, azimuth angles, and background noise; and (3) \textbf{Field recordings} using mobile recorders to capture authentic environmental ambiance. The speech content includes voice signals processed by multiple configurations of neural voice enhancement algorithms alongside unprocessed baseline recordings. For each binaural clip, 5 human subjective ratings were collected via a crowdsourced listening test using the structured evaluation template detailed in Section 2. Ground-truth ratings were obtained across 9 distinct attributes of the perceptual and spatial experience, as summarized in Table~\ref{tab:attributes_need}.

\begin{figure}[t]
  \centering
  \includegraphics[width=\linewidth]{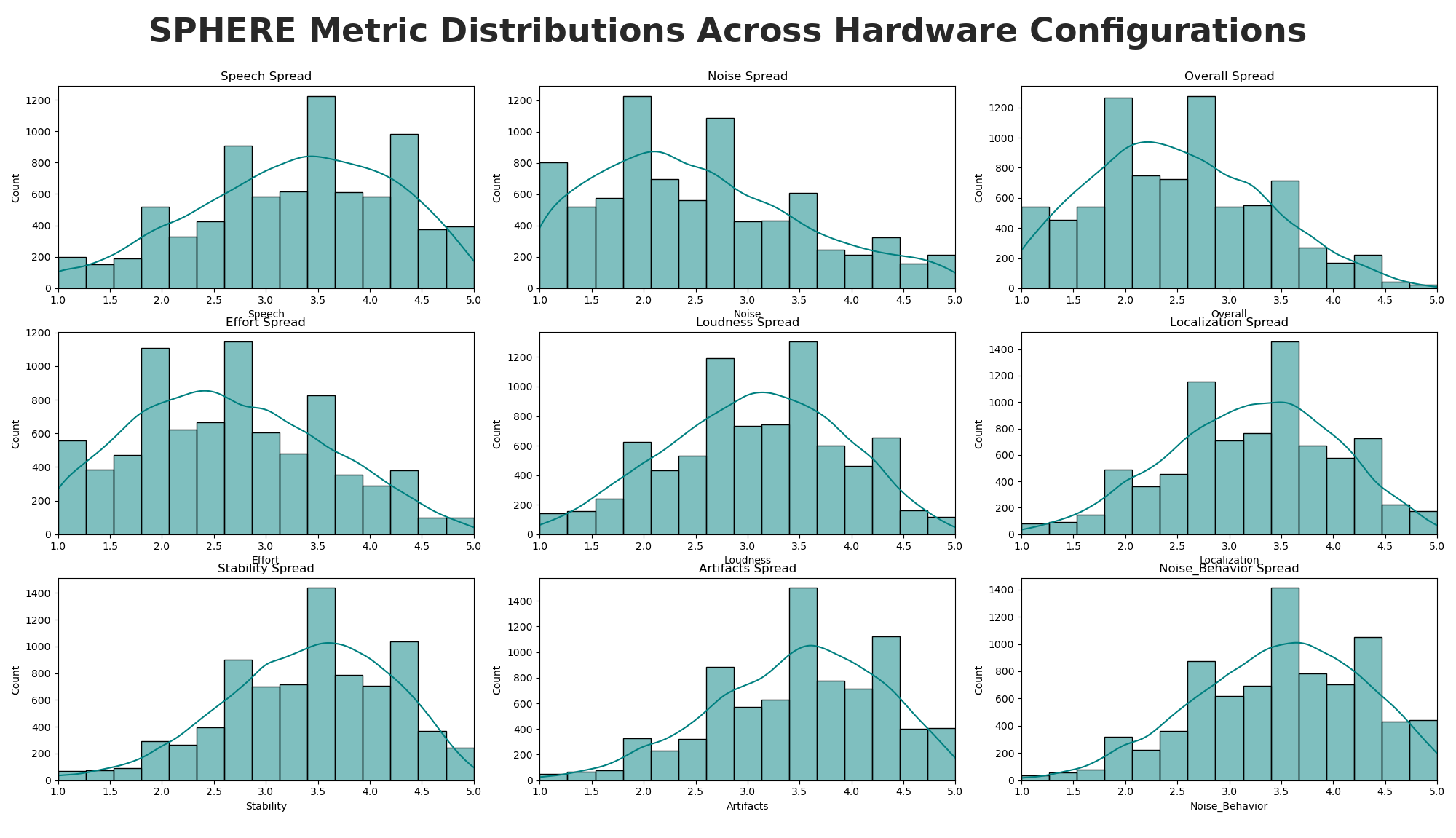}
  \caption{Distribution of human subjective Mean Opinion Scores (MOS) across the 9 evaluated attributes of the binaural speech dataset.}
  \label{fig:metric_distributions}
  \vspace{-0.3cm}
\end{figure}

\begin{figure}[t]
  \centering
  \includegraphics[width=0.9\linewidth]{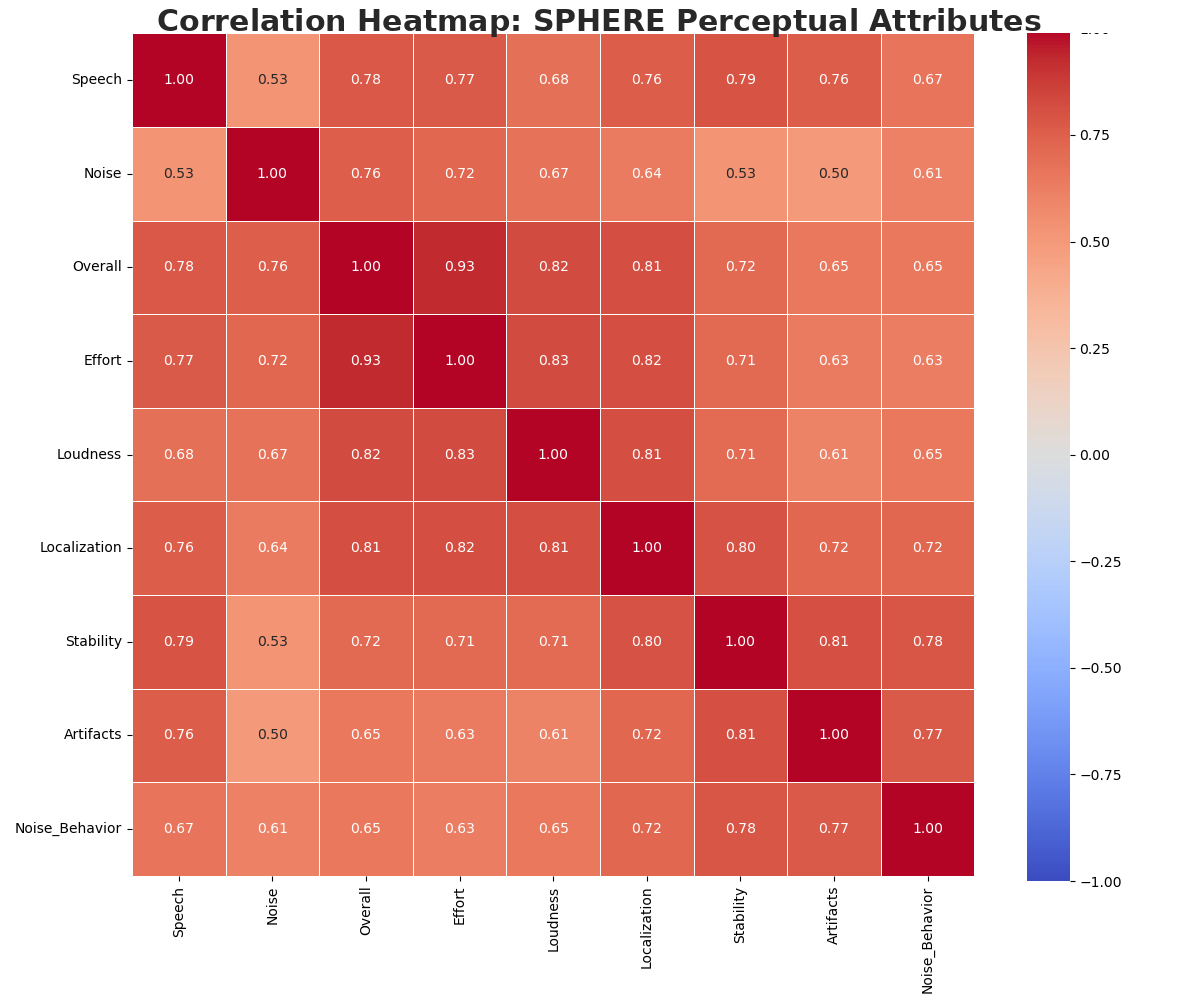}
  \caption{Pearson correlation matrix between the 9 subjective spatial audio metrics, illustrating high redundancy between overall quality and listening effort, and moderate correlation between speech and noise quality.}
  \label{fig:metric_correlation}
  \vspace{-0.3cm}
\end{figure}

The distribution of the mean opinion scores (MOS) of 5 ratings across all 9 dimensions is illustrated in Fig. \ref{fig:metric_distributions}, showing a comprehensive quality spectrum spanning the full range of 1.0 to 5.0. Fig. \ref{fig:metric_correlation} shows the correlation matrix between these metrics. We observe a very high correlation ($0.93$) between overall quality and listening effort, and a high correlation (0.81--0.82) between overall quality, spatial localization, and loudness comfort, demonstrating that spatial attributes are highly integrated into the overall human perceptual rating of binaural speech.

\subsection{Two-Stage Data Augmentation Pipeline}
To train deep networks effectively and prevent overfitting, we implement a two-stage data augmentation pipeline using Apache Beam. The pipeline branches the source data into a pristine dataset and an augmented training corpus: (1) \textbf{Sliding Window Chunking}, where a 10-second window is slid across each 20-second source clip with a 2-second stride, extracting 6 chunks per file and propagating parent MOS labels under a weak supervision assumption. While overall quality remains stationary across 20 seconds, transient localized defects introduce label noise for unaffected chunks, explaining lower correlations on defect metrics; and (2) \textbf{Perceptually Invariant Transformations}, applying random time shifting (up to $\pm 50$~ms), gain adjustments (up to $\pm 2.5$~dB), and polarity inversion preserving MOS validity. The unmodified first chunk (0--10~s) forms the pristine dataset, while all sliding chunks and their invariant augmented variants form the augmented training corpus.

\subsection{No-Leak Splitting Protocol}
To ensure a rigorous evaluation and prevent data leakage across partitions, we enforce a strict speaker- and session-level splitting protocol. The partitioning is performed on the unique clean base audio filenames before any chunking or augmentation transformations are applied. This guarantees that all augmented variations (simulated rooms, spatial positions, gains, and window crops) of a given source audio file are mapped strictly to the same partition, preventing any overlap of speaker characteristics or environmental acoustic contents between the training and validation sets. The base clean filename is resolved for every record by stripping all augmentation and chunk suffixes. Shuffling and splitting (80\% training, 20\% validation) are performed directly on this list of unique base filenames. All subsequent augmented variations of a clean audio file are mapped strictly to the same partition. Prior to feature extraction, waveforms are normalized to $[-1.0, 1.0]$ by scaling the 16-bit integer representation by $32768.0$. While the training set is fully augmented, we apply deterministic under-sampling to the validation split to keep the evaluation size manageable while preserving diversity. For each base file in the validation split, we keep the pristine chunk, and randomly select only 2 sliding window chunks (out of 5 possible) and 2 gain steps (out of 4).

\begin{table*}[t]
\centering
\caption{Performance comparison across the 9 perceptual dimensions on the leak-free validation set ($N=2,119$ clips). Attribute-level Pearson Correlation Coefficients (Clip PCC, $r_p$) and 9-attribute model averages across all evaluation metrics are reported. For the monaural DNSMOS baseline~\cite{dnsmos_p835}, standard per-channel score averaging ($(f(x_L)+f(x_R))/2$) is reported on its 3 supported P.835 dimensions (9-attribute averages are omitted as DNSMOS cannot predict 6 of the 9 dimensions). Best results in \textbf{bold}.}
\label{tab:results}
\setlength{\tabcolsep}{5pt}
\resizebox{\textwidth}{!}{
\begin{tabular}{l|c|cccccc}
\toprule
\textbf{Perceptual Metric} & \textbf{DNSMOS \cite{dnsmos_p835}} & \textbf{Non-Siamese MEL} & \textbf{Siamese MEL} & \textbf{Non-Siamese STFT} & \textbf{Siamese STFT} & \textbf{Non-Siamese MEL + SF} & \textbf{Siamese MEL + SF} \\
\midrule
Overall Quality (MOS-OVRL)  & 0.284 & 0.759 & 0.766 & 0.764 & 0.750 & \textbf{0.773} & \textbf{0.773} \\
Listening Effort            & ---   & 0.776 & 0.781 & 0.781 & 0.769 & \textbf{0.784} & 0.783 \\
Background Noise (MOS-BAK)  & 0.460 & 0.789 & 0.800 & 0.779 & 0.785 & \textbf{0.807} & 0.801 \\
Speech Quality (MOS-SIG)    & 0.054 & \textbf{0.669} & 0.663 & 0.664 & 0.636 & 0.656 & \textbf{0.669} \\
Loudness Comfort            & ---   & 0.755 & 0.745 & 0.747 & 0.736 & \textbf{0.759} & 0.756 \\
Sound Localization          & ---   & \textbf{0.710} & 0.697 & 0.697 & 0.659 & 0.706 & 0.700 \\
Spatial Stability           & ---   & 0.637 & 0.627 & \textbf{0.644} & 0.603 & 0.643 & 0.636 \\
Distracting Artifacts       & ---   & \textbf{0.632} & 0.607 & 0.625 & 0.573 & 0.618 & 0.611 \\
Background Noise Behavior   & ---   & 0.593 & 0.568 & 0.604 & 0.540 & \textbf{0.611} & 0.589 \\
\midrule
Average Clip PCC ($r_p \uparrow$)        & --- & 0.702 & 0.695 & 0.700 & 0.673 & \textbf{0.706} & 0.702 \\
Average Clip SRCC ($r_s \uparrow$)       & --- & 0.660 & 0.652 & 0.655 & 0.630 & 0.659 & \textbf{0.660} \\
Average Clip RMSE ($\downarrow$)         & --- & 0.744 & 0.733 & 0.742 & 0.752 & 0.734 & \textbf{0.726} \\
Average Clip MAE ($\downarrow$)          & --- & 0.599 & 0.589 & 0.600 & 0.604 & 0.591 & \textbf{0.582} \\
Average System PCC ($r_{sys} \uparrow$)  & --- & 0.715 & \textbf{0.770} & 0.752 & 0.688 & 0.728 & 0.730 \\
Average Quality Tier PCC ($r_{tier} \uparrow$) & --- & \textbf{0.994} & \textbf{0.994} & \textbf{0.994} & 0.992 & \textbf{0.994} & 0.993 \\
\bottomrule
\end{tabular}
}
\vspace{-0.35cm}
\end{table*}

The resulting split yields a training set of 433,396 records (1.394~TB stored in ArrayRecord format across 50 shards) and a multi-dimensional validation set of 12,554 records ($\sim$40~GB) categorized into standard, gain-augmented, and sliding-window variants.

\section{Model Architectures}
\label{sec:architectures}

To establish the feasibility of predicting the 9 perceptual MOS metrics automatically, we evaluate two modeling paradigms: a single-branch (Non-Siamese) network and a two-branch (Siamese) weight-shared network. Both models employ a hybrid CNN-RNN architecture designed to capture spectral, temporal, and spatial cues from 10-second stereo audio segments.

\subsection{Stereo Baseline Model (Non-Siamese)}

\begin{figure}[t]
\centering
\resizebox{0.82\columnwidth}{!}{
\begin{tikzpicture}[
    node distance=0.25cm and 0.8cm,
    block/.style={draw, rectangle, minimum height=0.35cm, minimum width=1.8cm, align=center, fill=blue!10, rounded corners=1pt, font=\scriptsize},
    shared/.style={draw, rectangle, minimum height=0.35cm, minimum width=1.8cm, align=center, fill=green!10, dashed, rounded corners=1pt, font=\scriptsize},
    arrow/.style={->, >=stealth, thick}
]
    \node (in_l) [block] {Left Audio\\(10s @ 16kHz)};
    \node (spec_l) [block, below=of in_l] {STFT / Mel\\Spectrogram};
    \node (cnn_l) [shared, below=of spec_l] {2D CNN Encoder\\(Shared $\theta$)};
    
    \node (in_r) [block, right=0.8cm of in_l] {Right Audio\\(10s @ 16kHz)};
    \node (spec_r) [block, below=of in_r] {STFT / Mel\\Spectrogram};
    \node (cnn_r) [shared, below=of spec_r] {2D CNN Encoder\\(Shared $\theta$)};
    
    \node (concat) [block, below=0.4cm of $(cnn_l.south)!0.5!(cnn_r.south)$] {Fusion Layer\\(Concatenate $L$, $R$, $|L-R|$, and $L \times R$)};
    
    \node (bilstm) [block, below=of concat] {Stacked BiLSTM\\(2 layers, 128 units)};
    \node (pool) [block, below=of bilstm] {Global Average Pooling};
    
    \node (heads) [block, below=of pool] {9 Separate Heads\\(Dense + Dropout)};
    \node (out) [block, below=of heads] {9 Perceptual MOS Outputs\\(Concatenated vector $[B, 9]$)};
    
    \draw [arrow] (in_l) -- (spec_l);
    \draw [arrow] (spec_l) -- (cnn_l);
    \draw [arrow] (in_r) -- (spec_r);
    \draw [arrow] (spec_r) -- (cnn_r);
    
    \draw [arrow] (cnn_l.south) -- (concat);
    \draw [arrow] (cnn_r.south) -- (concat);
    \draw [arrow] (concat) -- (bilstm);
    \draw [arrow] (bilstm) -- (pool);
    \draw [arrow] (pool) -- (heads);
    \draw [arrow] (heads) -- (out);
    
    \path (cnn_l) -- node[above, font=\scriptsize\itshape] {Shared $\theta$} (cnn_r);
    \draw[<->, dashed, thick, color=gray] (cnn_l) -- (cnn_r);
    
\end{tikzpicture}
}
\caption{Detailed architecture of the two-branch Siamese network evaluated in SPHERE.}
\label{fig:siamese_architecture}
\vspace{-0.3cm}
\end{figure}
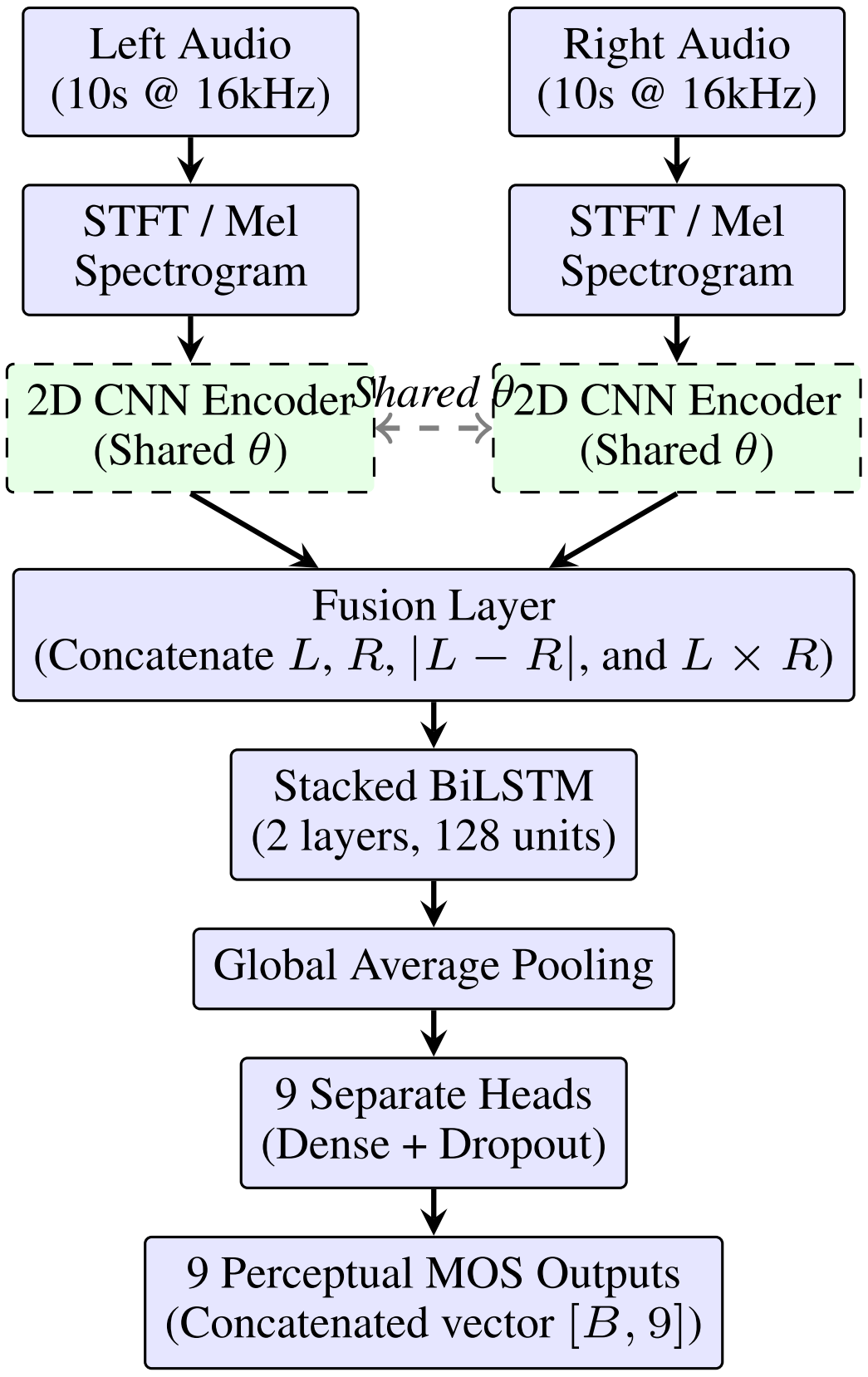
The Non-Siamese baseline processes stereo audio by first extracting dual-channel features (either log-magnitude Mel-spectrograms or Short-Time Fourier Transform (STFT) magnitudes) from the left and right channels. These features are stacked along the channel dimension, yielding a tensor of shape $[B, T, F, 2]$, where $B$ is batch size, $T$ is time frames, and $F$ is frequency bins. This 2-channel representation is fed into a 2D CNN encoder consisting of five convolutional blocks with filter sizes (32, 32, 64, 64, 128) and batch normalization, followed by max-pooling layers.

The output is reshaped to $[B, T, D]$ ($D=1024$ for Mel). A temporal trunk of 2 stacked BiLSTM layers (128 units) captures bidirectional temporal context. Global Average Pooling yields a fixed-size representation of shape $[B, 256]$. For multi-task regression, this is routed to 9 independent heads (64-unit Dense with ReLU, dropout 0.5, and 1-unit linear projection), totaling $\sim$321k parameters.

\subsection{Symmetric Two-Branch Siamese Architecture}
Binaural perception relies on Interaural Time Differences (ITD) and Level Differences (ILD). To explore symmetric processing, we evaluate a Siamese architecture \cite{mittag2020full, morgadoNIPS18}. As shown in Fig.~\ref{fig:siamese_architecture}, left and right channels are processed by identical, weight-shared 2D CNN encoders ($\text{Shared }\theta$). For STFT, complex STFT features $[B, T, F, 2] $ stack real and imaginary parts.

The CNN outputs are merged in a fusion layer $\mathbf{h}_{\text{fused}} = \left[ \mathbf{h}_L \,;\, \mathbf{h}_R \,;\, |\mathbf{h}_L - \mathbf{h}_R| \,;\, \mathbf{h}_L \odot \mathbf{h}_R \right]$ and projected to match the baseline BiLSTM input dimension, maintaining comparable capacity ($\sim$330k parameters).

\subsection{Spatial Feature Fusion (MEL + SF)}
To evaluate self-supervised spatial embeddings, the \textbf{MEL + SF} configuration augments the Mel baseline with pretrained PhaseCoder \cite{dementyev2026phasecoder} representations. PhaseCoder extracts a 256-D embedding capturing microphone-geometry agnostic interaural phase structures. This vector is z-score normalized, projected via a 64-unit Dense layer with SiLU, and concatenated with the 256-D pooled BiLSTM trunk ($[B, 320]$) before routing to the 9 prediction heads ($\sim$353k params).

\section{Experimental Evaluation}
\label{sec:experiments}

\subsection{Training Setup}
Models are trained with Adam (learning rate $2\times10^{-4}$, batch size 128, TPU) with L2 regularization ($10^{-4}$). The objective is Weighted Mean Squared Error (WMSE) with equal attribute weights ($w_k = 1.0, \forall k$): $\mathcal{L} = (1/N) \sum_n \sum_{k=1}^9 w_k (\hat{y}_{n,k} - y_{n,k})^2$. Checkpoints are validated every 500 steps, retaining the model with maximum validation PCC.

\subsection{Evaluation Metrics and Methodology}
Models are benchmarked across five complementary statistical criteria:
(1) \textbf{Clip PCC ($r_p$)}: Pearson correlation coefficient evaluating linear alignment across all $N=2,119$ individual validation clips.
(2) \textbf{Clip SRCC ($r_s$)}: Spearman rank correlation coefficient assessing monotonic ranking accuracy invariant to non-linear score compression.
(3) \textbf{Absolute Error (RMSE, MAE)}: Root Mean Square Error and Mean Absolute Error quantifying absolute deviations on the 1.0--5.0 MOS scale.
(4) \textbf{System-Level PCC ($r_{sys}$)}: Evaluates algorithm benchmarking fidelity per ITU-T P.808/P.835. For each enhancement condition $s \in \mathcal{S}$ ($N=6$ systems, e.g., unprocessed audio, classic beamforming/DSP, and 4 neural voice enhancement baselines), clip ratings and predictions are averaged ($\bar{y}_s, \bar{\hat{y}}_s$). System PCC is the Pearson correlation across condition means, filtering per-clip idiosyncratic acoustic variance.
(5) \textbf{Quality Tier PCC ($r_{tier}$)}: Evaluates global calibration linearity across 5 discrete subjective quality intervals ($[1, 2]$, $(2, 3]$, $(3, 3.5]$, $(3.5, 4]$, and $(4, 5]$) by averaging predictions and ratings per tier.

\textbf{Clip-Level Correlation \& Error}: All configurations show robust clip-level tracking (Avg. PCC $\approx 0.70$, SRCC $\approx 0.66$, RMSE $\approx 0.73$, MAE $\approx 0.58$). Augmenting Mel spectrograms with PhaseCoder embeddings (\textbf{MEL + SF}) achieves the highest clip-level correlation (Avg. PCC = \textbf{0.706}) and lowest error (Avg. RMSE = \textbf{0.726}, MAE = \textbf{0.582} in Siamese MEL + SF), while improving rank correlation on Overall Quality (attribute-level SRCC = \textbf{0.708} vs. $0.687$). Furthermore, Quality Tier PCC exceeds \textbf{0.99} across all models, confirming monotonic calibration across the degradation spectrum.

\textbf{System-Level Ranking vs. Clip-Level Noise (Why Siamese?)}: While unconstrained 2D filters in Non-Siamese models show a marginal advantage on raw per-clip PCC ($0.706$ vs. $0.702$), the \textbf{Siamese architecture decisively outperforms Non-Siamese at the System Level} (Avg. System PCC of \textbf{0.770} vs. $0.715$). Weight-shared Siamese encoders enforce physical symmetry and interaural consistency (ITD/ILD), yielding dramatic gains on spatial attributes: \textbf{Sound Localization} jumps from $0.793$ to \textbf{0.910}, and \textbf{Spatial Stability} jumps from $0.786$ to \textbf{0.867}. Siamese models also excel on Background Noise ($0.715$ vs. $0.665$) and Distracting Artifacts ($0.822$ vs. $0.768$), confirming that symmetric inductive bias is essential for benchmarking spatial audio systems.
\textbf{Comparison with Monaural Baseline (DNSMOS)}: To quantify the need for native binaural modeling, we evaluated the open-source ITU-T P.835 DNSMOS predictor~\cite{dnsmos_p835} (16~kHz) on the same validation set ($N=2,119$). Since DNSMOS accepts only single-channel inputs, we evaluated both time-domain downmixing ($x_{\text{mono}} = (x_L + x_R)/2$) and standard per-channel score averaging ($\hat{y} = (f(x_L) + f(x_R))/2$). Acoustic downmixing introduces severe phase cancellation and comb-filtering due to interaural time differences (ITDs) and HRTFs ($\text{MOS-SIG } r_p = 0.039$, $\text{MOS-OVRL } r_p = 0.256$). While per-channel score averaging (reported in Table~\ref{tab:results}) avoids comb-filtering and improves correlation ($\text{MOS-OVRL } r_p = 0.284$, $\text{MOS-BAK } r_p = 0.460$, $\text{MOS-SIG } r_p = 0.054$), it still lags behind SPHERE by $+0.347$ to $+0.615$ PCC points because monaural models cannot capture binaural unmasking or interaural spatial cues and cannot predict listening effort or spatial attributes.
\section{Conclusion}

We introduced SPHERE, a 9-attribute subjective evaluation framework for binaural speech quality, listening effort, and spatial fidelity. Evaluated on a crowdsourced dataset with strict hardware pre-filtering and no-leak splitting, baseline neural architectures reliably predict all 9 attributes (up to 0.706 clip PCC, 0.770 system PCC, robust SRCC, and low RMSE/MAE). Crucially, system aggregation demonstrates that symmetric Siamese processing is vital for spatial fidelity (0.910 localization PCC), providing a foundational benchmark for next-generation hearables.

\bibliographystyle{IEEEbib}
\bibliography{refs}

@article{doclo2010binaural,
  title={Binaural noise reduction algorithms for hearing aids},
  author={Doclo, Simon and Kellermann, Walter and Makino, Shoji and Nordholm, Sven E},
  journal={IEEE Signal Processing Magazine},
  volume={27},
  number={1},
  pages={11--30},
  year={2010},
  publisher={IEEE}
}

@inproceedings{visqol,
  title={ViSQOLAudio: An objective audio quality metric for low bitrate codecs},
  author={Hines, Andrew and Gillen, Eoin and Kelly, Damien and Skoglund, Jan and Kokaram, Anil and Harte, Naomi},
  booktitle={The Journal of the Acoustical Society of America},
  volume={137},
  number={4},
  pages={2298--2298},
  year={2015},
  organization={Acoustical Society of America}
}

@inproceedings{sesqa,
  title={SESQA: Semi-supervised learning for speech quality assessment},
  author={Serr{\`a}, Joan and Pons, Jordi and Santiago, Pascual},
  booktitle={ICASSP 2021-2021 IEEE International Conference on Acoustics, Speech and Signal Processing (ICASSP)},
  pages={381--385},
  year={2021},
  organization={IEEE}
}

@article{schadler2020binaural,
  title={Binaural speech intelligibility benefits with directional microphones and noise reduction algorithms},
  author={Sch{\"a}dler, Marc Ren{\'e} and Warzybok, Anna and Kollmeier, Birger},
  journal={Trends in Hearing},
  volume={24},
  pages={2331216520938220},
  year={2020},
  publisher={SAGE Publications Sage UK: London, England}
}

@inproceedings{valin2024ar,
  title={Audio Processing for Augmented Reality and Smart Glasses},
  author={Valin, Jean-Marc and others},
  booktitle={IEEE International Conference on Acoustics, Speech and Signal Processing (ICASSP)},
  year={2023}
}

@INPROCEEDINGS{pesq,
  author={Rix, A.W. and Beerends, J.G. and Hollier, M.P. and Hekstra, A.P.},
  booktitle={2001 IEEE International Conference on Acoustics, Speech, and Signal Processing. Proceedings (Cat. No.01CH37221)}, 
  title={Perceptual evaluation of speech quality (PESQ)-a new method for speech quality assessment of telephone networks and codecs}, 
  year={2001},
  volume={2},
  number={},
  pages={749-752 vol.2},
  doi={10.1109/ICASSP.2001.941023}}

@article{polqa,
  title={Perceptual Objective Listening Quality Assessment (POLQA), The Third Generation ITU-T Standard for End-to-End Speech Quality Measurement Part I--Temporal Alignment},
  author={Beerends, John G. and Schmidmer, Christian and Berger, Jan and Oome, Maarten and Ptasinski, Rajmund and Hoogendoorn, Cornelis},
  journal={Journal of the Audio Engineering Society},
  volume={61},
  number={6},
  pages={366--374},
  year={2013}
}

@inproceedings{ambiqual,
  title={{AMBIQUAL}: Towards a Quality Metric for Ambisonics},
  author={Narbutt, Miroslaw and Skoglund, Jan and Allen, Andrew and Chinen, Michael},
  booktitle={Proc. AES International Conference on Audio for Virtual and Augmented Reality},
  year={2020}
}

@techreport{peaq,
  author={{ITU-R}},
  title={Method for objective measurements of perceived audio quality},
  type={Recommendation},
  number={BS.1387-2},
  institution={International Telecommunication Union},
  address={Geneva, Switzerland},
  year={2014}
}

@techreport{ITUTP810,
    author      = {{International Telecommunication Union}},
    title       = {P.810: Modulated noise reference unit (MNRU)},
    institution = {International Telecommunication Union},
    type        = {Recommendation},
    number      = {ITU-T P.810},
    year        = {2023},
    url        = {https://itu.int}
}

@techreport{ITUTP830,
  author      = {{ITU-T}},
  title     = {Subjective Performance Assessment of Telephone-Band and Wideband Digital Codecs},
  type      = {Recommendation},
  number    = {P.830},
  institution = {International Telecommunication Union},
  address   = {Geneva, Switzerland},
  year      = {1996},
  month     = aug
}

@techreport{ITUT_P811,
  author      = {{International Telecommunication Union}},
  title       = {{ITU-T Recommendation P.811: Subjective test methodology for evaluating speech oriented stereo communication systems over headphones}},
  institution = {International Telecommunication Union},
  type        = {Recommendation},
  number      = {P.811},
  year        = {2019},
  address     = {Geneva, Switzerland},
  url         = {https://itu.int}
}

@inproceedings{NISQA,
	doi = {10.21437/interspeech.2021-299},  
	year = 2021,
	month = {Aug},
	publisher = {{ISCA}},
	author = {Gabriel Mittag and Babak Naderi and Assmaa Chehadi and Sebastian M{\"o}ller},
	title = {{{NISQA}: A Deep {CNN}-Self-Attention Model for Multidimensional Speech Quality Prediction with Crowdsourced Datasets}},
	booktitle = {Interspeech 2021}
}

@misc{p835,
      title={Subjective test methodology for evaluating speech communication systems that include noise suppression algorithm.}, 
      author={ITU-T Recommendation P.835},
      year={2003},
      publisher = {{International Telecommunication Union, Geneva}
},
}

@misc{p808,
      title={Subjective evaluation of speech quality with a crowdsourcing approach.}, 
      author={ITU-T Recommendation P.808},
      year={2018},
      publisher = {{International Telecommunication Union, Geneva}
},
}

@inproceedings{dnsmos_p835,
  title={{{DNSMOS} {P.835}: A non-intrusive perceptual objective speech quality metric to evaluate noise suppressors}},
  author={Reddy, Chandan KA and Gopal, Vishak and Cutler, Ross},
  booktitle={ICASSP 2022-2022 IEEE International Conference on Acoustics, Speech and Signal Processing (ICASSP)},
  pages={886--890},
  year={2022},
  organization={IEEE}
}

@inproceedings{mittag2020full,
  title={Full-Reference Speech Quality Estimation with Attentional Siamese Neural Networks},
  author={Mittag, Gabriel and M{\"o}ller, Sebastian},
  booktitle={Proc. ICASSP},
  pages={346--350},
  year={2020}
}

@inproceedings{morgadoNIPS18,
  title={Self-Supervised Generation of Spatial Audio for 360° Video},
  author={Morgado, Pedro and Vasconcelos, Nuno and Langlois, Timothy and Wang, Oliver},
  booktitle={Proc. NeurIPS},
  year={2018}
}

@article{dementyev2026phasecoder,
  title={{PhaseCoder}: Microphone geometry-agnostic spatial audio understanding for multimodal {LLMs}},
  author={Dementyev, Artem and Zulfikar, Wazeer and Hersek, Sinan and Getreuer, Pascal and Kumar, Anurag and Kumar, Vivek},
  journal={Proc. ICML},
  year={2026}
}

\end{document}